\documentclass[12pt]{article}
\usepackage{amsmath,,calrsfs}
\usepackage{amsfonts}
\usepackage{amssymb}
\usepackage{amscd}
\usepackage{bbm}
\usepackage{fancybox}
\usepackage{cite}
\usepackage{amsmath,amsfonts,amsbsy}
\usepackage{pstricks,pst-node}
\usepackage[small,bf,hang]{caption2}
\usepackage{graphicx}
\usepackage{epsfig}
\usepackage{psfrag}
\usepackage{comment}

\usepackage{float}

\psset{unit=1.3cm,linewidth=.5pt,radius=.2}  

\usepackage{multirow}                     
\usepackage{float}                          
\usepackage{lscape}                         
\usepackage{bm}

\newcommand{\cD}{{\cal D}}

\newcommand{\cL}{{\cal L}}
\newcommand{\cM}{{\cal M}}

\newcommand{\cP}{{\cal P}}

\newcommand{\bb}{\bar\beta}

\newcommand{\beq}{\begin{equation}}
\newcommand{\eeq}{\end{equation}}
\newcommand{\bi}{\begin{itemize}}
\newcommand{\ei}{\end{itemize}}
\newcommand{\bt}{\begin{tabular}}
\newcommand{\et}{\end{tabular}}
\newcommand{\bc}{\begin{center}}
\newcommand{\ec}{\end{center}}

\newcommand{\be}{\begin{equation}}
\newcommand{\ee}{\end{equation}}
\newcommand{\bea}{\begin{eqnarray}}
\newcommand{\eea}{\end{eqnarray}}
\newcommand{\ba}{\begin{array}}
\newcommand{\ea}{\end{array}}

\def\bbox{{\,\lower0.9pt\vbox{\hrule \hbox{\vrule height 0.2 cm
\hskip 0.2 cm \vrule height 0.2 cm}\hrule}\,}}
\newcommand{\dsl}{\pa \kern-0.5em /}

\font\mybb=msbm10 at 10pt
\def\bb#1{\hbox{\mybb#1}}

\def\bR {\bb{R}}

\def\tr{{\rm tr}}

\makeatletter \@addtoreset{equation}{section} \makeatother

\def\slashchar#1{\setbox0=\hbox{$#1$}           
   \dimen0=\wd0                                 
   \setbox1=\hbox{/} \dimen1=\wd1               
   \ifdim\dimen0>\dimen1                        
      \rlap{\hbox to \dimen0{\hfil/\hfil}}      
      #1                                        
   \else                                        
      \rlap{\hbox to \dimen1{\hfil$#1$\hfil}}   
      /                                         
   \fi}

\begin{document}

\begin{titlepage}
\begin{center}

\hfill  RUG-CTN-2014-67, \ DAMTP-2014-20, \ MIT-CTP 4544

\vskip 1.5cm

{\Large \bf Minimal Massive 3D Gravity}

\vskip 1cm

{\bf Eric Bergshoeff\,${}^1$, Olaf Hohm\,${}^2$, Wout Merbis\,${}^1$,}
\vskip .2truecm

{\bf Alasdair J. Routh\,${}^3$ and Paul K. Townsend\,${}^3$} \\

\vskip 25pt

{\em $^1$ \hskip -.1truecm
\em Centre for Theoretical Physics,
University of Groningen, \\ Nijenborgh 4, 9747 AG Groningen, The Netherlands\vskip 5pt }

{email: {\tt e.a.bergshoeff@rug.nl; w.merbis@rug.nl}} \\

\vskip .4truecm

{\em $^2$ \hskip -.1truecm
\em  Center for Theoretical Physics, Massachusetts Institute for Technology,\\ Cambridge, MA 02139, USA\vskip 5pt }

{email: {\tt  ohohm@mit.edu}} \\

\vskip .4truecm

{\em $^3$ \hskip -.1truecm
\em  Department of Applied Mathematics and Theoretical Physics,\\ Centre for Mathematical Sciences, University of Cambridge,\\
Wilberforce Road, Cambridge, CB3 0WA, U.K.\vskip 5pt }

{email: {\tt A.J.Routh@damtp.cam.ac.uk; pkt10@damtp.cam.ac.uk}} \\

\end{center}

\vskip 0.5cm

\begin{center} {\bf ABSTRACT}\\[3ex]
\end{center}
We present an alternative to  Topologically Massive Gravity (TMG) with the same ``minimal'' bulk properties; i.e.~a single local degree of freedom that is realized as a  massive graviton in linearization about an anti-de Sitter (AdS) vacuum. However,  in contrast to TMG,  the new ``minimal massive gravity'' has both a positive energy graviton and  positive central charges for the asymptotic AdS-boundary conformal algebra.

\end{titlepage}
\section{Introduction}

Topologically Massive Gravity (TMG) is a parity-violating  extension of three-dimensional (3D) General Relativity (GR) that propagates, on
linearization about a maximally symmetric vacuum, a single  massive spin-2 mode \cite{Deser:1981wh}. Its action augments the Einstein-Hilbert action (plus ``cosmological'' term with cosmological parameter $\Lambda_0$)  by a  Chern-Simons action for  the Levi-Civita affine connection 1-form $\Gamma$. Omitting a positive  factor proportional to 
the inverse of the 3D Newton constant, which has dimensions of inverse mass, the TMG action is\footnote{We use a ``mostly plus'' metric signature convention.}
\begin{equation}\label{TMGaction}
S_{\rm TMG} \ = \  \int \!d^3x\, \sqrt{-\det g}\,  (\sigma R -2\Lambda_0)+ \frac{1}{2\mu}\int \! \tr \left\{ \Gamma d\Gamma + \frac{1}{3} \Gamma^3\right\}  \,, 
\end{equation}
where $\mu$ is a mass parameter,  and $\sigma$ is a sign (plus for GR  but  minus for TMG if  we insist on positive energy for the spin-2 mode).  The  TMG field equation derived from this action is
\begin{equation}\label{TMGfield}
\frac{1}{\mu} C_{\mu\nu} + \sigma  G_{\mu\nu}  + \Lambda_0 g_{\mu\nu} =0 \, ,
\end{equation}
where $G_{\mu\nu}$ is the Einstein tensor,  and $C_{\mu\nu}$  the (symmetric traceless) Cotton tensor, defined as
\begin{equation}
C_{\mu\nu} \equiv \frac{1}{\sqrt{-\det g}} \, \varepsilon_\mu{}^{\tau\rho} D_\tau S_{\rho\nu}\, , \qquad S_{\mu\nu} \equiv R_{\mu\nu} - \frac{1}{4}g_{\mu\nu} R\, .
\end{equation}
Here, $D$ is the covariant derivative defined with $\Gamma$, and $S_{\mu\nu}$  the 3D Schouten tensor.  As  $G_{\mu\nu}= -\Lambda g_{\mu\nu}$
for maximally symmetric vacua with cosmological constant $\Lambda$, and since $C_{\mu\nu}$ is zero in such vacua, the relation between $\Lambda_0$ and $\Lambda$ for TMG is $\Lambda= \sigma\Lambda_0$. 

TMG is a minimal theory of massive 3D gravity in the sense that a field propagating a single spin-2 mode in a Minkowski vacuum defines a unitary irrep of the 3D Poincar\'e group.
An obvious question is whether TMG could be the semi-classical approximation to some  3D quantum gravity theory; in particular, it is natural to wonder whether there might be a holographically dual conformal field theory (CFT) on the boundary  of an AdS$_3$ vacuum of TMG with cosmological constant  $\Lambda=-1/\ell^2$ for AdS$_3$ radius $\ell$. 
The trouble with this idea is that the central charge of such a CFT, computed in a semi-classical approximation, is negative whenever the  bulk spin-$2$ mode has positive energy, implying a non-unitary CFT.  A closely related problem is that the  Ba\~nados-Teitelboim-Zanelli (BTZ)  black hole solutions  (which exist for any 3D gravity theory with an AdS$_3$ vacuum)  have negative mass whenever the energy of bulk graviton modes is positive.  It was suggested in \cite{Li:2008dq} that this problem might be circumvented by first choosing $\sigma=1$, to ensure positive mass BTZ black holes, and then tuning the dimensionless parameter $\mu\ell$  to a critical point  at which the  bulk mode  is absent, and the boundary CFT is chiral, for sufficiently strong boundary conditions. However, another bulk mode appears at the critical point \cite{Carlip:2008jk}, and it was soon realized that this is  a chirality-violating ``logarithmic''  mode that  is compatible with consistent AdS$_3$ boundary conditions \cite{Grumiller:2008qz}, implying a non-unitary ``logarithmic'' boundary  CFT. 

Until recently, a similar state of affairs held for the parity-preserving New Massive Gravity (NMG) \cite{Bergshoeff:2009hq,Bergshoeff:2009aq},  which is a 4th-order extension of 3D
GR that propagates a parity doublet of spin-2 modes (and is therefore minimal with respect to the product of the Poincar\'e group with parity). This too suffers from the defect 
that the central charge of the boundary CFT dual is negative whenever the bulk spin-2 modes have positive energy, and tuning to critical points again leads only to non-unitary
``logarithmic''  boundary CFTs.  Extensions of TMG or NMG with  higher powers  of curvature have been discussed in the literature \cite{Sinha:2010ai,Paulos:2010ke} but the 
``bulk vs boundary clash'' persists. Other attempts to evade this conflict (e.g. \cite{Banados:2009it}) typically introduce, as a by-product,    the Boulware-Deser ghost 
\cite{Boulware:1973my} (which is  invisible in a linearized approximation). However,  the recently constructed ``Zwei Dreibein Gravity'' (ZDG) shows that there is a viable alternative to NMG \cite{Bergshoeff:2013xma,Bergshoeff:2014bia}. 

At present  there is no known alternative to TMG that resolves the  ``bulk vs boundary clash''  while preserving  the minimal bulk properties. The main purpose of this paper is to present such an alternative.  Since it has the same minimal local structure as TMG (and also for another reason to be explained later)  we shall call it ``Minimal Massive Gravity'' (MMG);
the essential difference is that the field equation of MMG includes the additional, curvature-squared, symmetric tensor\footnote{The overall sign in the definition of this tensor 
corrects that appearing in the published version; the sign given here is what is needed for the validity of the expression for $\gamma$ of eq. (\ref{translate}).}
\begin{eqnarray}\label{Jtensor}
J_{\mu\nu} &=&  \frac{1}{2\det g}\ \varepsilon_\mu{}^{\rho\sigma} \varepsilon_\nu{}^{\tau\eta} S_{\rho\tau}S_{\sigma\eta}  \nonumber \\
 &=&   R_\mu{}^\rho R_{\rho\nu} - \frac{3}{4} RR_{\mu\nu} - \frac{1}{2} g_{\mu\nu} \left(R^{\rho\sigma}R_{\rho\sigma} - \frac{5}{8} R^2\right)  \, .
\end{eqnarray}
In other words, the MMG field equation is
\begin{equation}\label{modTMG}
\frac{1}{\mu} C_{\mu\nu} + \bar\sigma  G_{\mu\nu}  + \bar\Lambda_0 g_{\mu\nu} =- \frac{\gamma}{\mu^2} J_{\mu\nu} \, ,
\end{equation}
where $\gamma$ is some non-zero dimensionless constant; we have replaced $\sigma$ by $\bar\sigma$ since it is no longer obvious why it should be just a sign, and we have replaced $\Lambda_0$ by $\bar\Lambda_0$ since we  should not expect it to equal the cosmological parameter when $\bar\sigma=1$. 

Using the Bianchi identities satisfied by the Einstein and Cotton tensors, we see that consistency requires
$D_\mu J^{\mu\nu} =0$, but a direct computation shows that
\begin{equation}\label{Jid}
\sqrt{-\det g}\, D_\mu J^{\mu\nu} =    \varepsilon^{\nu\rho\sigma} S_\rho{}^\tau C_{\sigma\tau}\, .
\end{equation}
The right hand side is not identically zero, but it is only required to be zero as a consequence of the field equation (\ref{modTMG}); using this
to eliminate the Cotton tensor, we find that
\begin{equation}\label{onshellid}
\sqrt{-\det g}\, D_\mu J^{\mu\nu}=  \frac{\gamma}{\mu} \varepsilon^{\nu\rho\sigma} S_\rho{}^\tau J_{\tau\sigma} \equiv 0\, .
\end{equation}
The identity holds because the tensor $J$ can be written as 
\begin{equation}\label{Jeq}
J_{\mu\nu} =  S_\mu{}^\rho S_{\rho\nu} - S S_{\mu\nu} - \frac{1}{2}g_{\mu\nu}\left(S^{\rho\sigma}S_{\rho\sigma}-S^2\right)\, , 
\end{equation}
where $S\equiv g^{\mu\nu}S_{\mu\nu}$. This tells us that ``$\varepsilon SJ$'' is a linear combination  of terms of the form $\varepsilon S^n$ for $n=1,2,3$, which are all zero because $S^n$ is symmetric.  Thus, remarkably, the modified field equation (\ref{modTMG}) is consistent. 

The manner in which the MMG field equation (\ref{modTMG}) evades inconsistency is novel; we are not aware of any other example in which
consistency is achieved in this way.  We elaborate on the implications of this novelty  in the final section of this paper. One implication  is that  the MMG field equation (for the metric alone) cannot be obtained from any conventional higher-curvature  modification of the TMG action (\ref{TMGaction}).  Nevertheless,  there is a very  simple action with auxiliary 
fields that yields  precisely the equation (\ref{modTMG}), and the required auxiliary fields are those  already present in the dreibein formulation of TMG \cite{Grumiller:2008pr,Carlip:2008qh}.  These auxiliary fields can be eliminated from the field equations derived from the MMG action, leaving only the MMG equation (\ref{modTMG}). However, in contrast to the 
usual situation for auxiliary fields, back-substitution into the action is not legitimate  (for $\gamma\ne0$)  so the MMG action with auxiliary fields does not imply the existence of one without them.

These results would be no more than curiosities if the proposed modification of TMG were to change the local structure in an unacceptable way. A first indication
that this will not happen is that the $J$ tensor does not contribute to linearisation about a Minkowski vacuum, but this is a rather weak test. It  would seem to require
a miracle for this property to be maintained for other vacua, leaving aside the Boulware-Deser ghost.  Nevertheless, the miracle occurs, as we explain in detail in this
paper using Hamiltonian methods. Furthermore,  not  only is the local structure of MMG exactly that of TMG, but the freedom allowed in MMG permits a resolution
of the ``bulk vs boundary clash'', as we show by a computation of the algebra of the asymptotic conformal group in an asymptotically AdS$_3$
spacetime.

\section{Chern-Simons-like formulation}
\label{sec:CSLM}

We were led to the MMG model of massive gravity by considering possible modifications of TMG in the context of its formulation as a ``Chern-Simons-like'' model of
massive gravity \cite{Hohm:2012vh}. The idea is to consider an action that is the integral of a Lagrangian 3-form constructed as  a sum of exterior products of  $N$ ``flavours'' of Lorentz-vector valued 1-form fields $\{a^r; r=1,2 \dots, N \}$, which should include a dreibein $e^a$ and a dualised  spin-connection $\omega^a$.  The Lagrangian 3-form is assumed to take the form\footnote{The exterior product of forms is implicit.}
\begin{equation}\label{CSlike0}
L_{CSL} = \frac{1}{2}g_{rs}\eta_{ab} a^{r a} d a^{s b} + \frac{1}{6}f_{rst}\epsilon_{abc} a^{r a} a^{s b} a^{t c}\, ,
\end{equation}
where $g_{rs}$ is a constant flavour-space metric, assumed invertible, and $f_{rst}$ is a constant totally symmetric flavour tensor; these should be chosen such that the action
is locally Lorentz invariant.  The $N=2$ case is GR viewed as a Chern-Simons (CS) theory, but the generic $N\ge3$ case, which includes TMG \cite{Grumiller:2008pr,Carlip:2008qh},  has local degrees of freedom and is therefore only ``CS-like''.  The $N=4$ case  includes NMG and ZDG as well as parity-violating extensions of them \cite{Bergshoeff:2014bia}.  

Using a 3D vector algebra notation for Lorentz indices, we may rewrite  (\ref{CSlike0}) in a less cluttered notation as
\begin{equation}\label{CSlike}
L_{CSL} = \frac{1}{2}g_{rs} a^r \cdot d a^s + \frac{1}{6}f_{rst} a^r \cdot a^s \times a^t\, .
\end{equation}
For present purposes, we focus on the $N=3$ case  with $(e,\omega,h)$ as the three Lorentz vector-valued 1-form fields. From the first two of these we
can construct the locally Lorentz covariant torsion and curvature 2-forms
\begin{equation}
T(\omega)= d e +\omega \times e\, , \qquad R(\omega)= d\omega+ \frac{1}{2}\omega \times \omega\, .
\end{equation}
The $h$ field has the same  parity (odd) and dimension as $\omega$ and it appears in the TMG action as a Lagrange multiplier for the zero torsion constraint. The Lagrangian 3-form is
\begin{equation}
L_{TMG} = -\sigma e \cdot R + \frac{\Lambda_0}{6}e\cdot e\times e + h \cdot T(\omega) + \frac{1}{2\mu}\left(\omega \cdot d \omega + \frac{1}{3}\omega\cdot \omega \times \omega\right) \,,
\end{equation}
where  $\Lambda_0$ is a  cosmological parameter with dimensions of mass-squared, and $\sigma$ a sign. The last term, with a factor of $1/\mu$, is the ``Lorentz Chern-Simons'' (LCS) term, but here for the independent dual spin-connection $\omega$. The mass parameter  $\mu$ could have either sign since a parity transformation would effectively flip the sign of $\mu$. In the limit that $|\mu|\to\infty$ the TMG action reduces (for $\sigma=1$) to an action for 3D GR.  

Because  the 3D Newton constant has  dimensions of inverse mass, the Lagrangian 3-form should have dimension of mass-squared, as it does if we assign zero dimension to $e$ and dimensions of mass to  both $\omega$ and $h$. With these assignments of parity and dimension,  and given the requirement of  local Lorentz invariance, the TMG Lagrangian 3-form is  almost unique, up to field redefinitions, if we suppose that parity is broken only by the LCS term. However, there is one further parity-even term that could be included; this is an ``$ehh$'' term. This leads us to consider the following
one-parameter family of ``Minimal Massive Gravity'' (MMG) Lagrangian 3-forms
\begin{equation}\label{Letmg}
L_{MMG} = L_{TMG} + \frac{\alpha}{2} e \cdot h \times h\, ,
\end{equation}
where $\alpha$ is a dimensionless parameter.  In the absence of the parity-violating LCS term, the new ``ehh'' term is inoccuous; it leads only to an alternative action for 3D GR. 
However, when combined with the LCS term it leads, as we shall now show,  to a modification of TMG that is equivalent to the one described  in tensor form in the introduction (with the constant $\gamma$ being a function of $\alpha$). 

The field equations derived from the Lagrangian 3-form \eqref{Letmg} are
\begin{equation}
\begin{split} \label{etmgeom}
0 &= T(\omega) + \alpha e \times h \,, \\
0 &= R(\omega) + \mu e\times h  - \sigma \mu T(\omega) \,, \\
0 &= - \sigma R(\omega)+ \frac{\Lambda_0}{2} e\times e + D(\omega) h + \frac{\alpha}{2} h\times h \,.
\end{split}
\end{equation}
An equivalent set of equations is
\begin{eqnarray}\label{inOm}
0 &=& T(\Omega)\, , \nonumber\\
0 &=& R(\Omega) + \frac{\alpha\Lambda_0}{2} e\times e + \mu\left(1+ \sigma\alpha\right)^2 e\times h\, , \\
0 &=& D(\Omega)h - \frac{\alpha}{2} h\times h + \sigma\mu \left(1+\sigma\alpha\right) e\times h  + \frac{\Lambda_0}{2}\, e\times e\, , \nonumber
\end{eqnarray}
where the new dual spin-connection 1-form is
\begin{equation}
 \Omega = \omega + \alpha h\, .
 \end{equation}
In what follows we shall assume that the dreibein is invertible.
Using the identities
\begin{equation}
D(\Omega)T(\Omega) \equiv R(\Omega)\times e\, , \qquad D(\Omega)R(\Omega) \equiv 0\, ,
\end{equation}
and assuming that 
\begin{equation}\label{alphacon}
1+ \sigma\alpha\ne 0\, ,
\end{equation}
one finds that the field equations imply
\begin{equation}\label{second}
0= e\cdot h \equiv e^a h^b \eta_{ab} \, .
\end{equation}

The first of  equations (\ref{inOm}) implies that $\Omega= \Omega(e)$,  the usual torsion free spin-connection, which can be traded for the Levi-Civita affine connection.  
The second of equations (\ref{inOm}) can be solved for $h$:
\begin{equation}\label{hsol}
h_{\mu\nu} \equiv h_\mu{}^a e_\nu{}^b \eta_{ab} =  - \frac{1}{\mu\left(1+\alpha\sigma\right)^2} \left[S_{\mu\nu} + \frac{\alpha\Lambda_0}{2} g_{\mu\nu}\right]\, ,
\end{equation}
where $S_{\mu\nu}$ is the 3D Schouten tensor. Since this is a symmetric tensor, we learn that $h_{\mu\nu}$ is symmetric; this is precisely the content of (\ref{second}).
At this point, we have used the first two of equations (\ref{inOm}) to solve for $\Omega$ and $h$ in terms of $e$. Because of local Lorentz invariance, back-substitution 
into the action will produce an action for the metric alone. However, this back-substitution is not legitimate when $\alpha\ne0$ because the equations used are {\it not} equivalent, jointly, to the two equations found by varying the action with respect to $\Omega$ and $h$; to get them one needs (if $\alpha\ne0$) to use the  $e$-equation in addition to the $\Omega$ and $h$ equations. 

Although it is not legitimate to back-substitute into the action, it is legitimate to substitute the expressions for $\Omega$ and $h$ into the third equation of  (\ref{inOm})  to arrive at an equation for the metric\footnote{The situation is similar for  ZDG; the equations allow the elimination of one dreibein in terms of the other one but back-substitution into the action is not legitimate. One can still substitute into the ZDG equations to get a field equation for only one dreibein but  it  involves an infinite series that must be constructed order by order  \cite{Bergshoeff:2014eca}.}.  Doing so, we  find that this equation is equivalent to the MMG tensor equation (\ref{modTMG}) with
coefficients
\begin{equation}\label{translate}
\bar\sigma = \sigma +\alpha\left[1+ \frac{\alpha\Lambda_0/\mu^2}{2\left(1+\sigma\alpha\right)^2}\right]\, , \qquad \gamma= -\frac{\alpha}{\left(1+ \sigma\alpha\right)^2} \, ,
\end{equation}
and\footnote{Recall  that $\Lambda_0$ is the cosmological parameter occurring in the Lagrangian,  $\bar\Lambda_0$ the constant coefficient of $g_{\mu\nu}$ in the field equation
and $\Lambda$  the cosmological constant of a vacuum solution of these equations. For TMG  $\Lambda= \sigma\bar\Lambda_0 =\sigma\Lambda_0$  but for  other 3D gravity models, including MMG, the relation between these parameters is more complicated.}
\begin{equation}
\bar\Lambda_0 = \Lambda_0 \left[1+ \sigma\alpha - \frac{\alpha^3 \Lambda_0/\mu^2}{4\left(1+\sigma\alpha\right)^2}\right]\, .
\end{equation}

We have now found an action, with auxiliary fields,  that yields the MMG equation (\ref{modTMG}). In its CS-like form it is a very simple and natural modification of the TMG action. However, {\it it cannot be reduced to an action for the metric alone by elimination of the auxiliary fields}. This result  accords with our  earlier observation that the MMG equation 
cannot be obtained from an action for the metric alone. 

We shall  later give a detailed proof that MMG has the same number of local degrees of freedom as TMG,  but the essence of the analysis is  as follows.
Starting from the CS-like action,  a time/space split leads directly to a constrained Hamiltonian system with the time components  acting  as Lagrange multipliers for $9$ primary constraints. The  $18$ space components  are the canonical variables. By construction,  $6$ of the primary constraints are first-class, generating diffeomorphisms and local Lorentz transformations, which leaves $3$ primary constraints. If any of these are first-class there will be additional gauge invariances, so any such model will be exceptional.  For example, if  all $3$ remaining primary constraints are first-class then the CS-like theory is actually a CS theory with no local degrees of freedom. This possibility is realized when  $\alpha\sigma=-1$ because $L_{MMG}$ is then the sum of a CS 3-form for $(e,\Omega)$ (with gauge group depending on $\Lambda_0$) and an $Sl(2;\bR)$ CS 3-form for $\omega$; we exclude this case by imposing the restriction (\ref{alphacon}).

One can go systematically through the other possibilities for the general $3$-flavour model (allowing for the possibility of secondary constraints) to show that the dimension of the  physical phase space, per space point, must be either $0$ or $2$. As we shall see, only the latter possibility is consistent with a linearized analysis of MMG (assuming $1+\sigma\alpha\ne0$)  so the generic case  is the one of relevance here.  This is the case for which the three remaining primary constraints form a second-class set together  with one secondary constraint, the space component of   (\ref{second}). This yields a total of $6$ first-class and $4$ second-class constraints, implying a physical phase space with dimension per space point of $(18 - 2\times 6 - 1\times 4) = 2$, corresponding to one local degree of freedom, exactly as for TMG.

\section{Linearized analysis}

We now look for maximally symmetric vacuum solutions of the MMG equations in the form (\ref{inOm}). These are solutions for which
\be
R(\Omega) = \frac{1}{2}\Lambda\,  e \times e\, , \qquad \left( \ \Rightarrow\quad S_{\mu\nu}= \frac{1}{2}\Lambda g_{\mu\nu}\right)\, ,
\ee
where $\Lambda$ is the cosmological constant.  Substitution into the equation of motion \eqref{modTMG} yields
\begin{equation}\label{ha}
h = C\mu \, e \, , \qquad \sigma\Lambda/\mu^2= \Lambda_0/\mu^2 - \alpha\left(1+\sigma\alpha\right)C^2 \, ,
\end{equation}
for dimensionless constant
\begin{equation}
C= - \frac{\left(\Lambda +\alpha\Lambda_0\right)/\mu^2}{2\left(1+\sigma\alpha\right)^2} = - \frac{\Lambda}{2\mu^2} + \mathcal{O}(\alpha)\, .
\end{equation}

When $\alpha=0$ we have $\sigma\Lambda=\Lambda_0$; otherwise we have a quadratic equation for
$\Lambda$, for which the solution is
\begin{equation}\label{lambdasol}
\Lambda = -\alpha\Lambda_0 - \frac{2\mu^2\left(1+ \sigma\alpha\right)^3}{\sigma\alpha}
 \left[1 \mp \sqrt{1+ \frac{\alpha\Lambda_0/\mu^2}{\left(1+\sigma\alpha\right)^2}}\right]\, .
\end{equation}
To recover the $\alpha=0$ case in the $\alpha\to0$ limit, one must choose the upper sign. From now on, we will reserve the upper sign for the TMG branch, i.e.~the branch that contains the TMG model in the $\alpha \rightarrow 0$ limit. The
lower sign  denotes the non-TMG branch.

For real $\Lambda$ we must restrict $\Lambda_0$ such that
\begin{equation}\label{restrict}
\alpha\Lambda_0/\mu^2 + \left(1+\sigma\alpha\right)^2\ge 0\, .
\end{equation}

\subsection{Linear equations}

We now linearize about an AdS background, for which
\begin{equation}
\Lambda = - 1/\ell^2\, ,
\end{equation}
where $\ell $ is the AdS radius.  Let  $\bar{e}$ be the background dreibein, and $\bar{\omega}\equiv \Omega(\bar e)$ the background spin connection.  We expand about this background by writing
\begin{equation}
e = \bar{e} + k\,, \qquad \Omega = \bar{\omega} +  v\,, \qquad
h = C\mu \left( \bar{e} +  k \right) + p\,,
\end{equation}
where $(k,v,p)$ are perturbations; $k$ has even parity whereas $v$ and $p$ have odd parity. The expansion of $h$ breaks parity due to the presence of a term linear in $\mu$.
To first order in these perturbations, the field equations are\,\footnote{These equations break parity but only because of terms involving odd powers of the mass $\mu$.}
\begin{equation}\label{lin}
\begin{split}
& \bar{D} k + \bar{e}\times v = 0\,, \\
& \bar{D} v - \Lambda \, \bar{e}\times k = - \mu (1 + \alpha \sigma)^2 \bar{e}\times p \,, \\
& \bar{D} p + M \bar{e}\times p = 0\,,
\end{split}
\end{equation}
where $\bar D$ is the covariant exterior derivative for spin-connection $\bar\omega$, and 
\begin{equation}\label{mass}
M = \left[\sigma(1+ \sigma\alpha) - \alpha C\right]\mu =  \pm\, \sigma\mu \left(1+\sigma\alpha\right)\sqrt{1 + \frac{\alpha\Lambda_0}{\mu^2\left(1+\sigma\alpha\right)^2}} \, .
\end{equation}
The sign here is the same as the one appearing in (\ref{lambdasol});  in other words, the top sign allows an $\alpha\to0$ limit, whereas the bottom sign does not.  

Notice that   the condition (\ref{restrict}), required for reality of $\Lambda$, is equivalent to $M^2\ge0$.  Let us also record here, for future use, the identity
\begin{equation}\label{CM}
1-2C = \frac{(\ell M)^2-1}{(1+\sigma\alpha)^2 (\ell\mu)^2}\, .
\end{equation}

The integrability conditions of the equations (\ref{lin}) may be found by using  the fact that for any Lorentz-vector valued 1-form $a$, 
\begin{equation}
\bar D^2 a=  \frac{1}{2} \Lambda (\bar e\times \bar e) \times a =\Lambda \, \bar e \left(\bar e \cdot a\right)\, . 
\end{equation}
This leads to the conclusion that the equations (\ref{lin}) imply 
\begin{equation}
\bar e\cdot p =0\, . 
\end{equation}
Equivalently, these equations imply that the tensor field $p_{\mu\nu}\equiv p_\mu{}^a \bar e_\nu{}^b\eta_{ab}$ on AdS$_3$ is symmetric.

Provided that   $|\ell M|\ne1$ (equivalently, $2C \ne1$)  the  set of three first-order equations  (\ref{lin}) may be diagonalized. This is achieved by introducing the new variables 
$f_\pm^a$ defined by
\begin{equation}
\label{cov}
k^a = \ell(f^a_{+} + f^a_{-}) + \frac{1}{\mu\left(1 - 2C\right)}\,  p^a\,, \qquad v^a = f^a_{+} - f^a_{-} + \frac{M}{\mu\left(1 -2C\right)}\,  p^a\, .
\end{equation}
This leads to the three equations
\begin{equation}\label{three-eq}
\begin{split}
&\bar{D} f_+ +  \ell^{-1}  \bar{e}\times  f_+ = 0\,, \\
&\bar{D} f_- -  \ell^{-1}  \bar{e}\times  f_- = 0\, , \\
&\bar{D} p + M  \, \bar{e}\times p \ = 0\,.
\end{split}
\end{equation}
Parity now exchanges $f_+$ with $f_-$, so the equations for these fields are exchanged by parity. Taken together, these two equations preserve parity.
The equation for $p$ breaks parity, as expected because $M\propto\mu$; this is the AdS$_3$ version of the ``self-dual'' equation for a single massive 
spin-2 mode \cite{Aragone:1986hm}.

For any Lorentz vector-valued one-form field $a$ the first-order equation 
\begin{equation}\label{first-order}
\bar D a + m\,  \bar e\times a =0
\end{equation} 
is equivalent, given that $\bar e\cdot a=0$ and hence that $a_{\mu\nu}$ is symmetric\footnote{This condition is a consequence of (\ref{first-order}) when $|\ell m|\ne1$, and when $\ell m=\pm 1$ it may be imposed as a gauge condition.},  to the equation
\begin{equation}
\cD[(\ell m)^{-1}]_\mu{}^\nu a_{\nu\rho} =0 \, , \qquad 
{\cal D}[\eta]_\mu{}^\nu \equiv \ell^{-1} \delta_\mu{}^\nu - \frac{\eta}{\det \bar e}\,  \epsilon_\mu{}^{\tau\nu}{} {\bar D}_\tau \, , 
\end{equation}
combined with the condition that the symmetric tensor $a$ is traceless.  Using this result, we may rewrite the equations (\ref{three-eq}) in tensor form as 
\begin{equation}\label{another-three}
\cD[1] f_+=0\, , \qquad \cD[-1]f_- =0\, , \qquad  \cD[(\ell M)^{-1}] p=0
\end{equation}
for symmetric traceless tensors $(f_\pm,p)$. More generally, without assuming that $|\ell M|\ne1$, we may use the equations (\ref{lin})  to eliminate $p$ and $v$ and thus obtain the following third-order equation 
\begin{equation}\label{third-order}
(\cD\left[\left(\ell M\right)^{-1}\right]  \cD\left[-1\right]\cD\left[+1\right])_\mu{}^\nu k_{\nu\rho}= 0\, , 
\end{equation}
where the tensor $k$ is both symmetric and traceless. Evidently, this third-order equation is equivalent to the three first-order equations (\ref{another-three}) when $|\ell M|\ne1$. The $|\ell M|=1$ case yields the linearized equations of a ``critical''  MMG model with a ``logarithmic'' bulk mode; we shall not study this critical case here.

\subsection{Absence of tachyons}

The solutions of the first-order equation $\cD[\eta] a=0$, for symmetric traceless second-order tensor $a$,  define an irrep of the AdS$_3$ isometry group, which  is unitary
provided that $|\eta|\le1$, with $\eta=\pm1$ corresponding to the singleton irreps that have no bulk support (see e.g. \cite{Bergshoeff:2010iy} for a review).
It follows that of the three equations  (\ref{another-three})  only the one with $\eta= (\ell M)^{-1}$ propagates a bulk mode, which has spin-$2$ because $p$ is a 
symmetric traceless 2nd-order tensor. The condition for unitarity of the irrep defined by this equation is 
\begin{equation} \label{notach}
|\ell M| >1 \,  \qquad \left(\Leftrightarrow \quad 1-2C >0\right)\, . 
\end{equation}
An immediate consequence is that $M^2>0$, so the condition (\ref{restrict}) required for reality of $\Lambda$ will be satisfied. The more stringent condition (\ref{notach}) is equivalent to positivity of the graviton mass-squared. This is because 
\begin{equation}
\cD[(\ell M)^{-1}]k=0 \quad \Rightarrow \quad \cD[-(\ell M)^{-1}]\cD[(\ell M)^{-1}]k=0\, , 
\end{equation}
which is the Fierz-Pauli spin-2 field equation in AdS$_3$ for a spin-2 field $k$ with mass ${\cal M}$ given by
\begin{equation}
\ell^2{\cal M}^2 = \ell^2 M^2 -1 \, . 
\end{equation}
We may therefore interpret the condition (\ref{notach}) as a ``no-tachyon'' condition.

\subsection{Absence of ghosts}

We have still to determine the condition that the spin-2 bulk mode is not a ghost, but to do this we need to consider the quadratic action for the perturbations about the AdS$_3$ vacuum, not just the linearized field equations.  When the action (\ref{Letmg}) is expanded about the AdS$_3$ vacuum in terms of the one-form field fluctuations $(f_+,f_-,p)$ one finds, to quadratic order, an action that is the integral of  the Lagrangian $3$-form 
\begin{eqnarray}\label{quad}
L^{(2)} &=& \frac{\lambda_+}{\mu} \left[f_+ \cdot \bar D f_+  + \ell^{-1} \bar e\cdot f_+\times f_+ \right] +
\frac{\lambda_-}{\mu} \left[f_- \cdot \bar D f_-  - \ell^{-1} \bar e\cdot f_-\times f_- \right] \nonumber \\
&& + \frac{1}{2\mu\left(1-2C\right)}\ \left[p\cdot \bar D p + M \, \bar e\cdot p\times p\right]\, ,
\end{eqnarray}
where
\begin{equation}
\lambda_\pm = 1\mp \left(\sigma + \alpha C\right)\mu\ell\, .
\end{equation}
We shall see later that these coefficients  are directly related to the boundary central charges; in fact
\begin{equation}
c_\mp \ \propto  \ \mp \frac{\lambda_\pm}{\mu\ell}  =  \sigma \mp \frac{1}{\mu\ell} + \alpha C\, ,
\end{equation}
which agrees with the TMG result for $\alpha=0$.

Notice that
\begin{equation}
-\lambda_+\lambda_- =  \ell^2\mu^2\left(1-2C\right) >0\, ,
\end{equation}
where the final inequality is a consequence of the no-tachyon condition (\ref{notach}). This inequality implies that  $\lambda_+$ and $\lambda_-$ must have opposite sign, which  means that we can rescale the  $f_\pm$ fields to bring their contribution to $L^{(2)}$ into the form of the difference of two linearized $Sl(2;\bR)$ CS  3-forms.  Up to an overall sign,  this is the linearization of 3D GR with negative cosmological constant in its  CS-formulation, so the $f_\pm$ fields have no  local degrees of freedom, in agreement 
with the analysis of the previous subsection. 

Let us now focus on the term in $L^{(2)}$ that is quadratic in $p$; this is 
\begin{equation}\label{quadp}
L_p^{(2)} =  -AM \left(p\cdot \bar D p + M \, \bar e\cdot p\times p\right)\, , \qquad A= -\frac{1}{2M\mu\left(1-2C\right)}\, . 
\end{equation}
As already mentioned, the first-order field equation for $p$ implies the second-order Fierz-Pauli equation in  AdS$_3$. We now aim to examine this relationship at the level of the quadratic action because this will determine the condition for the  spin-2 mode 
propagated by the $p$-equation to have positive energy. To this end, consider the following action for 
 Lorentz-vector valued 1-forms  $q$ and $s$: 
\begin{equation}\label{Lfp}
L_{\rm F.P.} = - A \left\{q\cdot \bar{D}s+ \frac12 \bar{e}\cdot s \times s + \frac12M^2\,  \bar{e}\cdot q\times q\right\}\,, 
\end{equation}
where $A$ is a normalization constant. If we solve  the first-order field equation for $s$ and back-substitute then we find the following Lagrangian density for symmetric tensor $q_{\mu\nu} = q_{(\mu}{}^a \bar e_{\nu)}{}^b \eta_{ab}$ (the antisymmetric part drops out):\begin{equation}
\mathcal{L}_{\rm F.P.} = - A \left\{ q^{\mu\nu} \mathcal{G}_{\mu\nu}(q) + \frac12 \cM^2 \left( q_{\mu\nu} q^{\mu\nu} - q^2\right) \right\} \,,  
\qquad \left(q= \eta^{\mu\nu}q_{\mu\nu}\right)\, , 
\end{equation}
where  $\mathcal{G}_{\mu\nu}(q)$ is the linearized Einstein tensor and  ${\cal M}^2= M^2 +\Lambda$. This is the  Fierz-Pauli  Lagrangian density for a spin-2 field of mass 
${\cal M}$ in an AdS$_3$ background; see e.g. \cite{Bergshoeff:2009aq}, where the conventions used are the same as those used here.  From this result we  learn that the no-ghost condition is $A>0$.  Next, we diagonalise the Fierz-Pauli Lagrangian 3-form (\ref{Lfp}) by writing it in terms of the new Lorentz-vector one-form fields $q_\pm$ defined by 
\begin{equation}
q = q_+ + q_- \,, \qquad s = M q_+ - M q_-\, .
\end{equation}
We thus find that 
\begin{eqnarray}\label{Lmin}
L_{\rm F.P.} &=& - A M \left( q_+ \cdot \bar{D} q_+ + M \bar{e} \cdot q_+ \times q_+ \right) \nonumber \\
&& + \ A M \left( q_- \cdot \bar{D} q_- - M \bar{e} \cdot q_- \times q_- \right)\,.
\end{eqnarray}
The $q_\pm$ field propagates a single spin-2 mode of helicity $\pm 2$, and for both to have positive energy we require $A>0$. However, the two modes are exchanged by parity, which is a symmetry of the Fierz-Pauli action,  so if the helicity $\pm2$ mode has positive energy then so does the $\mp2$ mode. This means that $A>0$ is the condition for either helicity mode alone to have positive energy, the action for a single mode being obtained by setting either $q_+\equiv0$ or $q_-\equiv0$. In particular, we may set 
$q_-\equiv0$ to recover  the Lagrangian 3-form of (\ref{quadp}) with $q_+\equiv p$. We conclude that the no-ghost condition for this Lagrangian 3-form is 
$A>0$ with $A$ given in (\ref{quadp}); i.e. 
\begin{equation}
M\mu(1-2C)<0\, . 
\end{equation}

\subsection{Combined no-tachyon/no-ghost conditions}\label{subsec:combined}

The  no-tachyon and no-ghost conditions combined are equivalent to the two conditions
\begin{equation}
1-2C>0 \quad \& \quad  M/\mu <0\, . 
\end{equation}
These are equivalent to the two conditions
\begin{equation}
{\cal M}^2 >0 \quad \& \quad \pm \sigma(1+ \sigma\alpha) <0\, , 
\end{equation}
where the upper sign must be chosen if the AdS$_3$ vacuum is the one allowing an $\alpha\to0$ limit. The latter equation leads to the following possibilities:

\begin{enumerate}

\item {\it Top sign}:  $\sigma=-1$ and $0<\alpha<1$.  

\item {\it Top sign}:  $\sigma=-1$ and $\alpha=0$. This is the TMG case, for which the no-tachyon condition is $|\ell\mu|>1$. The fact that 
$\sigma=-1$ is the origin of the negative BTZ black hole mass and negative boundary central charge for TMG. 

\item {\it Top sign}:  $\sigma=-1$ and $\alpha<0$.

\item {\it Top sign}:  $\sigma=1$ and $\alpha<-1$.

\item {\it Bottom sign}: $\sigma=-1$ and $\alpha>1$.

\item {\it Bottom sign}: $\sigma=1$ and $-1<\alpha<0$.

\item {\it Bottom sign}: $\sigma=1$ and $\alpha >0$.

\end{enumerate}
We shall see later that only three of these seven possibilities survive when we add the condition of positive boundary central charges.

\section{Hamiltonian analysis}

We will  now analyse the constraint structure  to show that, quite generally, the Lagrangian 3-form (\ref{Letmg}) defines a model describing a single bulk degree of freedom.
Our analysis follows that of  \cite{Carlip:2008qh,Hohm:2012vh,Bergshoeff:2014bia}. In this analysis it is not necessary to consider any particular background but 
we will be interested in spacetimes that are asymptotic to an AdS vacuum, so we will pay attention to boundary terms. This will allow us  to  find
 the central charges in the sum of Virasoro algebras spanned by the conserved charges at the AdS boundary \cite{Brown:1986nw}.

 Starting from the generic  Lagrangian 3-form of (\ref{CSlike}), we separate the time and space components of the Lorentz-vector valued 1-forms by writing $
a= a_0 dt  +  a_i dx^i$.  This leads, on writing $ \varepsilon^{0ij}= \varepsilon^{ij}$, to the Lagrangian density
\begin{equation}\label{gentimedecomp}
\cL = - \frac12  \varepsilon^{ij} g_{rs}  a_{i}^r \cdot  \dot{a}_{j}^s + a_{0}^r \cdot \phi_r\, , 
\end{equation}
where the Lorentz vectors $a_0^r$ are Lagrange multipliers for primary constraints with Lorentz-vector constraint functions
\begin{equation}\label{constraints}
\phi_r = \varepsilon^{ij} \left(g_{rs}\,  \partial_i a_j^s + \frac12 f_{rst}\,   a_i^s \times  a_j^t \right)\,.
\end{equation}

We define the smeared constraint functionals $\phi[\xi]$ by integrating the constraint functions \eqref{constraints} against a  set of arbitrary Lorentz-vector fields $\xi^r$:
\begin{equation}
\varphi[\xi] = \int_{\Sigma} d^2x \; \xi_a^r \phi_r^a + Q[\xi] \,,
\end{equation}
where $\Sigma$ is a constant $t$ space-like hypersurface,  and $Q[\xi]$ is a boundary term, added to ensure  differentiability of the functionals $\varphi[\xi]$. 
In \cite{Bergshoeff:2014bia} it is shown that their Poisson brackets are given by
\begin{align} \label{gen_poissonbr}
\left\{ \varphi[\xi] , \varphi [\eta] \right\}_{\rm PB} = & \; \varphi[[\xi, \eta]] + \int_{\Sigma} d^2x \; \xi^r_a \eta^s_b \, \cP_{rs}^{ab}
\nonumber \\
& - \int_{\partial \Sigma} d\phi \; \xi^r \cdot  \left(g_{rs} \,  \partial_\phi\eta^s + f_{rst}\,   a_\phi {}^s \times  \eta^t   \right)\, ,
\end{align}
where $[\xi ,\eta]^t  = f^t{}_{rs}\, \xi^r \times  \eta^s$, and we have defined
\begin{equation}\label{Pmat_def}
\cP_{rs}^{ab} = f^t{}_{q[r} f_{s] pt}\,  \eta^{ab} \Delta^{pq}  +  2f^t{}_{r[s} f_{q]pt} (V^{ab})^{pq}\,,
\end{equation}
with
\begin{equation}
\Delta^{pq} = \varepsilon^{ij}  a_i^p \cdot  a_j^q \,, \qquad  V_{ab}^{pq}  =  \varepsilon^{ij} a^p_{i\, a} a^q_{j\, b} \, .
\end{equation}
The integration variable $\phi$ for the boundary term is the angular coordinate parametrizing $\partial\Sigma$, which is  the intersection of the boundary of AdS$_3$ with $\Sigma$. 

If some of the Lorentz vector valued one-forms are invertible then secondary constraints may arise from a set of integrability conditions, which can be obtained by acting on the field equations with an exterior derivative \cite{Bergshoeff:2014bia}.

\subsection{Local degrees of freedom of MMG}

Now we specialize to the MMG model defined by the Lagrangian three-form \eqref{Letmg}. As a consequence of the assumed invertibility of the dreibein, we find  that
\begin{equation}
\mu(1 +\sigma\alpha)^2\Delta^{eh} = 0\,.
\end{equation}
As we also assume $(1 + \sigma\alpha) \neq 0$,  this equation gives the additional (secondary) constraint $0= \Delta^{eh} \equiv \psi$. Taking account of this constraint,
we can omit the $\Delta^{eh}$ term from  the right hand side of \eqref{Pmat_def}, and there is then no  $\Delta^{pq}$ term. In the basis ($\omega,e,h $), the remaining term gives the $9\times 9$ matrix
\begin{equation}\label{matPB}
\cP = \mu(1 + \sigma\alpha)^2\left( \begin{array}{ccc}
0 & 0 & 0 \\
0 & - V^{hh}_{ab} & V^{he}_{ab} \\
0 & V^{eh}_{ab} & V^{ee}_{ab}
\end{array} \right) \, .
\end{equation}
We will also need the  Poisson brackets of the primary constraint functionals $\phi[\xi]$ with the one secondary constraint function $\psi$; this is
\begin{align}\label{PBSecCon}
\{ \varphi[\xi], \psi\}_{\rm PB} = \varepsilon^{ij} \big[ & D_{i} \xi^e\cdot  h_j - D_i \xi^h \cdot  e_j - \alpha\,  \xi^e\cdot h_i\times  h_j + \mu \sigma (1 + \alpha\sigma) \xi^e \cdot e_i \times h_j \nonumber \\
&  +  \left( \Lambda_0\, \xi^e+ \mu \sigma(1+\alpha\sigma) \xi^h \right) \cdot e_i \times e_j \big] \,,
\end{align}
where
\begin{equation}
D_i\xi^r = \partial_i \xi^r + \omega_i \times \xi^r\, .
\end{equation}
The $(9\times 9)$ matrix ${\cal P}$ has rank 2. When we combine this with the Poisson brackets of the secondary constraints, this increases the dimension of the matrix by one and the rank by two, since the brackets \eqref{PBSecCon} are independent of the column space defined by \eqref{matPB}. The final $(10\times10)$ matrix has rank 4, meaning that four constraints are second-class and the remaining six are first-class. The dimension of the  physical phase space, per space point, is then $3\times 6 -2\times 6 -4 =2$, implying a single bulk degree of freedom. This is consistent with the linear analysis of the last section, but we now know that this  is  a background independent property of the fully non-linear theory.

\section{Boundary central charge}

To extract the boundary central charge from the Poisson bracket algebra  \eqref{gen_poissonbr} it is sufficient to consider the AdS$_3$ boundary terms  for the two sets of  mutually commuting first-class constraints  \cite{Carlip:2008qh}. We must therefore identify the set of first-class constraints, which generate local Lorentz transformations and diffeomorphisms. To this end we define ({\it no sum on $r$})
\begin{equation}
\varphi_r[\xi^r] = \int_\Sigma\! d^2x\,  \xi^r\cdot\varphi_r + Q_r[\xi^r]\, , 
\end{equation}
where $Q_r$ is a surface term such that $\sum_r \varphi_r[\xi^r] = \varphi[\xi]$, and such that the functional derivative of $\varphi_r[\xi^r]$  is well-defined. 

It is easy to see that $\varphi_\omega[\chi]$ generates local Lorentz transformations,  with Lorentz-vector parameter $\chi$,  as  its Poisson brackets with all other constraints (including the secondary) vanish on the constraint surface.   The action, on canonical variables, of a diffeomorphism associated to a vector field $\zeta^\mu$ is generated by
\begin{equation}
\varphi_{\rm Diff}[\zeta] = \sum_r \varphi_r [ \zeta^\mu a_\mu^r] \, . 
\end{equation} 
Specifically, one finds that
\begin{equation}
\left\{ \varphi_{\rm Diff}[\zeta], a_i^r \right\}_{PB} = {\cal L}_\zeta a_i ^r +  {\rm equations \ of \ motion}\, , 
\end{equation}
where ${\cal L}_\zeta$ is the Lie derivative with respect to the vector field $\zeta^{\mu}$.

We now have a basis for the first class constraints.  We aim to find a new basis such that the Poisson-bracket algebra of first-class constraints becomes a direct sum of isomorphic 
algebras close to the AdS boundary.  To this end, we consider the linear combinations
\begin{eqnarray}
L_{\pm} [\xi] &=& \varphi_{\rm Diff}[\zeta] + \varphi_\omega[\zeta^\mu\left(a_\pm e_\mu - \omega_\mu)\right] \nonumber \\
&=& \varphi_e\left[\zeta^\mu e_\mu\right] + \varphi_h\left[\zeta^\mu h_\mu\right] + a_\pm \varphi_\omega\left[\zeta^\mu e_\mu\right]\, , 
\end{eqnarray}
for constants $a_\pm$. 
By making use of the general result \eqref{gen_poissonbr} and the fact that $h=\mu C e$ in the AdS vacuum, and hence close to the boundary of any asymptotically-AdS 
spacetime,  we find that
\begin{align}
\{L_+[\xi], L_-[\eta]\} = &  \left( 2 \alpha \mu C + a_+ + a_-\right) (\varphi_e[[\xi,\eta]]+ \mu C\, \varphi_h[[\xi,\eta]])  \nonumber \\
& + \left( 2\mu^2 (1+\alpha \sigma) C + a_+a_-\right) \varphi_{\omega}[[\xi,\eta]] + \ldots\,, 
\end{align}
where the dots denote boundary terms which will vanish after choosing suitable boundary conditions. The remainder of the right hand side vanishes when
\begin{equation}
a_{\pm} = - \alpha \mu C \left(1 \mp \sqrt{1+ \frac{2(1+\sigma\alpha)}{C\alpha^2}} \right)\,.
\end{equation}
Using this parametrization for $a_{\pm}$ and the identities
\begin{equation}
2\alpha\mu C + 2 a_{\pm} = \pm \frac{2}{\ell}\,, \qquad a_{\pm}^2 + 2\mu^2(1+\sigma\alpha)C = \pm \frac{2a_{\pm}}{\ell}\,,
\end{equation}
we find that 
\begin{equation}\begin{split}\label{PBLs}
\{ L_{\pm}[\xi], L_{\pm}[\eta] \}  = &  \pm \frac{2}{\ell} L_{\pm}[[\xi,\eta]]  \\
& \pm \frac{2}{\ell}\left(\sigma  \pm \frac{1}{\mu \ell} + \alpha C\right) \int_{\partial \Sigma}\! d\phi\, \xi \cdot \left[ \partial_{\phi} \eta +  \left( \bar{\omega}_{\phi}{} \pm \frac{1}{\ell} \bar{e}_{\phi}{}\right)\times \eta \right]\,. 
\end{split}
\end{equation}

After choosing suitable (Brown-Henneaux) boundary conditions and restricting the gauge transformations to those which preserve these boundary conditions, we can see that the boundary term in \eqref{PBLs} is responsible for a central extension in the asymptotic symmetry algebra of global charges. The expression obtained here is equivalent to the pure three dimensional gravity case \cite{Brown:1986nw}, with a modified expression for the central charge. After including the proper normalizations, we find that the asymptotic symmetry algebra consists of two copies of the Virasoro algebra with a central charge
\begin{equation}\label{app:centralcharge}
c_{\pm}  =  \frac{3 \ell}{2 G_3}\left(\sigma  \pm \frac{1}{\mu \ell} + \alpha C\right)\, ,
\end{equation}
where $G_3$ is the 3D Newton constant. 
Note that in the TMG limit $\alpha \to 0$ the central charges reduces to the TMG expressions. Unitarity of the dual CFT requires
these central charges to be positive.

\subsection{Positivity of the central charges}

We saw in subsection \ref{subsec:combined} that there are seven distinct choices of (i) AdS$_3$ vacuum branch (ii) sign of $\sigma$ and (iii) range of $\alpha$ for which the propagating spin-2 mode is neither a ghost nor a tachyon, and for each of these we require ${\cal M}^2>0$.  We shall now investigate the compatibility of these constraints with the requirement of positive central charges $c_\pm$; we shall see that only three of the seven cases survive.  For clarity, we summarize here the conditions that we wish to fulfil simultaneously:
\begin{align}\label{inequalities}
\text{No-tachyon\ \& No-ghost:} &&& 1-2C > 0\,, \quad \& \quad 
 \pm \sigma (1+\alpha\sigma) < 0 \, . \nonumber\\
\text{Positive\ central\ charges}: &&& \sigma  - \frac{1}{|\mu \ell|} + \alpha C >0\,. 
\end{align}
Here the $\pm$ in the no-ghost condition depends on the sign in \eqref{lambdasol} and hence differentiates the two branches of AdS vacua, which we shall now consider in turn. 
The  parameters are $\Lambda_0/\mu^2$ and $\alpha$, both of which may, a priori, be any real numbers, and $\mu\ell$, which we may assume to be positive without loss of generality. We find that the conditions (\ref{inequalities}) are satisfied simultaneously in the following three cases:
\begin{itemize}

\item {\it Top sign}: $\sigma=-1$,  $\alpha<0$ and 
\begin{equation} \label{paramTMG1}
\Lambda_0/\mu^2 = \frac{1}{\alpha ( \mu\ell)^2} + \frac{2(1-\alpha)^3}{\alpha^3} \left( 1+ \sqrt{1+\frac{\alpha^2/(\mu\ell)^2}{(1-\alpha)^2} } \right)\, . 
\end{equation}
As expected\footnote{Because we know that the conditions (\ref{inequalities}) are not simultaneously satisfied by TMG.}, this is singular at $\alpha=0$.  This is case (3)  of subsection \ref{subsec:combined}. In this case the dimensionless parameter $\gamma$ in the MMG field equation (\ref{modTMG}) 
is restricted to the range
\begin{equation}
0< \gamma \le  \frac{1}{4}\, . 
\end{equation}

\item {\it Top sign}: $\sigma=1$,  $\alpha<-1$ and 
\begin{equation} \label{paramTMG2}
 \Lambda_0/\mu^2 = \frac{1}{\alpha ( \mu\ell)^2} + \frac{2(1+\alpha)^3}{\alpha^3} \left( 1 - \sqrt{1+\frac{\alpha^2/(\mu\ell)^2}{(1+\alpha)^2} } \right) \, .
\end{equation}
This is case (4) of subsection \ref{subsec:combined}. In this case the dimensionless parameter $\gamma$ is restricted to the range
\begin{equation}
\gamma>0\, . 
\end{equation}

\item  {\it Bottom sign}: $\sigma = 1$, $-1< \alpha < 0$ and 
\begin{equation}
\Lambda_0/\mu^2 = \frac{1}{\alpha ( \mu\ell)^2} + \frac{2(1+\alpha)^3}{\alpha^3} \left( 1+ \sqrt{1+\frac{\alpha^2/(\mu\ell)^2}{(1+\alpha)^2} } \right) \,. 
\end{equation}
This is case (6) of subsection \ref{subsec:combined}.  In this case the dimensionless parameter $\gamma$ is again restricted to the range
\begin{equation}
\gamma>0\, . 
\end{equation}

\end{itemize}
Notice that $\alpha<0$, necessarily, and there are no ``bottom-sign'' cases with  $\sigma = -1$. Notice too that $\gamma>0$ in all cases. 

We remark that inversion of the formula (\ref{lambdasol}) giving $\Lambda$ as a function of $\Lambda_0$ yields
\begin{equation}\label{alLamzero}
\alpha\Lambda_0 = -\Lambda + \frac{2\mu^2\left(1+\sigma\alpha\right)^3}{\alpha^2} \left[1 \mp \sqrt{1- \frac{\alpha^2\Lambda/\mu^2}{\left(1+\sigma\alpha\right)^2}}\right]\, , 
\end{equation}
but there is no simple correlation of the sign with the vacuum branch sign of (\ref{lambdasol}). Given a choice of this branch, TMG (top sign)  or non-TMG (bottom sign), 
there is a definite value of $\Lambda$ for given $\Lambda_0$, and the sign in (\ref{alLamzero}) must then be chosen such that upon substitution for $\Lambda$ one recovers the given $\Lambda_0$. The signs in the above expressions for $\Lambda_0/\mu^2$ are such that this is the case, as may be checked by considering the first terms 
in an expansion in powers of $1/(\mu\ell)^2$ for $\mu\ell \gg 1$.

\section{Discussion}

We have presented a new, multi-parameter,  massive 3D gravity theory  that we have called ``Minimal Massive Gravity'' (MMG).  It is ``minimal'' in essentially two
different ways. 

One is that it shares with the well-known ``Topologically Massive Gravity'' (TMG) the property that it describes, when linearized about a flat or adS$_3$ vacuum, 
a single massive graviton mode. Like TMG, this mode is physical for some parameter range,  in the sense that it is neither a tachyon nor a ghost, and  there
are no other local degrees of freedom. TMG has a problem, however, when considered as a possible semi-classical limit of some quantum theory defined 
holographically via a dual CFT on an AdS$_3$ boundary. The asymptotic symmetry algebra  is a direct sum of two  Virasoro algebras \cite{Brown:1986nw} which must have positive central charges for unitarity of the CFT, and the parameters of TMG do not allow this condition to be satisfied while maintaining physical properties of the bulk mode.  In contrast, the one additional parameter of MMG allows a resolution of this ``bulk vs boundary'' clash; in fact we found three disjoint 
regions of parameter space  for which this is possible.  

It might appear from this summary that our resolution of the ``bulk vs boundary'' clash of TMG  has been achieved in a  rather obvious way (by the inclusion of extra terms in the action, leading to extra parameters and hence more freedom) and that MMG is really just a variant of TMG with more parameters. However, this is very far from being the case. 
As observed in the introduction, the MMG equation for the metric alone (after elimination of other, auxiliary, fields) cannot be found by variation of an action for the metric alone, 
so it does not correspond to any conventional extension of the TMG equation. In fact, MMG is {\it qualitatively} different from TMG, in various ways.  One of them leads to the conclusion that MMG is indeed ``minimal''  in another sense. 

Let us write the MMG equation (\ref{modTMG}) in the form $E_{\mu\nu}=0$, where
\begin{equation}
E_{\mu\nu} = \bar\Lambda_0 g_{\mu\nu}  + \bar\sigma  G_{\mu\nu}  + \frac{1}{\mu} C_{\mu\nu}  + \frac{\gamma}{\mu^2} J_{\mu\nu} \, . 
\end{equation}
Using the identity of (\ref{onshellid}), one finds that
\begin{equation}
\sqrt{-\det g}\, D_\mu E^{\mu\nu}  = \frac{\gamma}{\mu} \varepsilon^{\nu\rho\sigma} S_\rho{}^\tau E_{\tau\sigma}\, .
\end{equation}
Consistency of the MMG field equation requires the right hand side to be zero as a consequence of the MMG field equation, and this consistency condition is satisfied. 
However, let us now attempt to couple MMG to ``matter'' fields, which we suppose to have a (symmetric) stress tensor $T$ with the usual property that 
\begin{equation}\label{conserved}
D_\mu T^{\mu\nu}=0\, 
\end{equation}
as a consequence of the matter field equations.  Let us now consider (in some convenient units for  the 3D Newton constant) the equation 
\begin{equation}\label{E=T}
E_{\mu\nu} = T_{\mu\nu}\, . 
\end{equation}
In the case that $\gamma=0$, this is just the TMG field equation in the presence of matter with stress tensor $T$. Consistency of this equation requires that
\begin{equation}
D_\mu\left[E^{\mu\nu}-T^{\mu\nu}\right] =0\, , 
\end{equation}
and this is an identity for TMG, given (\ref{conserved}). However, for MMG ($\gamma\ne0$) we find, using (\ref{E=T}), that 
\begin{equation}\label{last}
\sqrt{-\det g}\, D_\mu\left[E^{\mu\nu}-T^{\mu\nu}\right] = \frac{\gamma}{\mu} \varepsilon^{\nu\rho\sigma}S_\rho{}^\tau T_{\tau\sigma}\, . 
\end{equation}
This is not zero unless the spacetime is an Einstein space, which is not required by the field equations\footnote{Alternatively, we could require the stress tensor $T$ to be a linear combination of the metric and Einstein tensors, but this would just change the coeficients in the source-free equation.}.  The difficulty here is that we are effectively assuming the standard minimal coupling of the metric to matter, such that variation of the matter action with respect to the metric yields, 
in a vacuum spacetime, the usual symmetric (Belinfante) stress tensor,  but there is no action for the metric alone to which this matter action could be added. It may be that some 
consistent matter couplings can be found by an extension of the CS-like action for MMG itself.  However, all that is clear at present  is  that the standard minimal coupling of gravity to matter is not possible.

We conclude with one further comment on MMG, which was actually our point of departure. As shown in  \cite{Bergshoeff:2014bia}, the Zwei-Dreibein Gravity model
of \cite{Bergshoeff:2013xma}, which resolves the ``bulk vs boundary clash'' for NMG,  has a parity-violating extension to a model that propagates two spin-2 modes of opposite 
3D-helicity with different masses. By sending the mass of one of the two modes to infinity,  one arrives at an alternative to TMG. This limit is rather subtle, but it leads to 
the MMG model described in this paper. This embedding of MMG into a model with additional degrees of freedom may be a useful source of further insight into its 
novel features. 

\bigskip 
{\bf Note added}: It has recently been shown that  the MMG equation can extended to include matter via a particular source tensor that is quadratic in the matter stress tensor  \cite{Arvanitakis:2014yja}. 
\bigskip

{\bf  Further note added}:  In our analysis of unitarity we assumed that the CFT central charges, as computed from the action, should both be positive. However, if they had  both been negative we could have
changed them to both positive by changing the sign of the action; this would also change the sign of the no-ghost condition, so it is really the combined no-ghost sign and the sign of the two central charges (assuming both have the same sign) that is physically relevant. Taking this possibility into account leads to new admissable ranges of  parameters for which $\gamma$ is negative rather than positive. However, a change in the sign of $\gamma$ can be compensated by a change in the sign of $\mu$ and of $\bar\sigma$ and $\bar\Lambda_0$, so the new admissable parameter ranges with $\gamma>0$ are physically equivalent to those with $\gamma>0$. 

\section*{Acknowledgements}
We are grateful to Hamid Afshar, Daniel Grumiller and Carl Turner for helpful discussions, and to Stanley Deser for correspondence.  A.~J.~R. thanks the members of the Centre for Theoretical Physics at the University of Groningen for their hospitality, and he acknowledges support from the UK Science and Technology Facilities Council. W.~M. is supported by the Dutch stichting voor Fundamenteel Onderzoek der Materie (FOM). O.~H. is supported by the U.S. Department of Energy (DoE) under the cooperative research agreement DE-FG02-05ER41360 and a DFG Heisenberg fellowship.

\providecommand{\href}[2]{#2}\begingroup\raggedright\endgroup


\end{document}